\documentclass[12pt]{iopart}

\usepackage{graphicx}

\begin{document}


\title{Local Quantum Critical Point and Non-Fermi Liquid Properties}

\author{Qimiao Si}

\address{Department of Physics \& Astronomy, Rice University,
Houston,TX 77005--1892, U.S.A.}

\date{\today}

\begin{abstract}
Quantum criticality provides a means to understand the apparent non-Fermi
liquid phenomena in correlated electron systems. How to properly describe
quantum critical points in electronic systems has however been poorly
understood. The issues have become particularly well-defined due to recent
experiments in heavy fermion metals, in which quantum critical points have 
been explicitly identified. In this paper, I summarize some recent
theoretical work on the subject, with an emphasis on the notion of 
``local quantum criticality''. I describe the microscopic work based on an
extended dynamical mean field theory, as well as Ginzburg-Landau arguments
for the robustness of the local quantum critical point beyond the
microscopics. 
I also present the consequences of this picture on
the inelastic neutron scattering, NMR, Fermi surface properties and Hall
coefficient, and compare them with the available experiments.
Some analogies
with the Mott transition phenomena are also
noted.
\end{abstract}



\section{Introduction}

One basic issue in correlated electron systems concerns
how electron-electron interactions lead to non-Fermi liquid behavior.
Proximity to a quantum critical point (QCP)
provides one mechanism. Fermi liquid theory 
for spatial dimensions higher than one is internally
consistent when the electron-electron interactions 
are treated perturbatively.
At a QCP, however, the effective interactions can become very
strong due to quantum critical fluctuations,
opening the door to a non-Fermi liquid critical state.

Quantum critical metals are of general interest in a variety of
strongly correlated electron systems, possibly also for
high temperature superconductors\cite{Tallon}.
The issues are however particularly well-defined 
in heavy fermion metals, for the simple reason that
QCPs have been explicitly identified.
Here, the transitions are typically between a paramagnetic metal
and an antiferromagnetic metal.
For instance, ${\rm CeCu_{6-y}Au_y}$ becomes magnetic
when the Au-doping reaches $y_c \approx 0.1$\cite{Lohneysen},
the stoichiometric ${\rm Yb Rh_2Si_2}$
is fortuitously 
sited 
($T_N \approx 70 mK$) 
close to its QCP\cite{Steglich},
and in the cases of ${\rm CePd_2Si_2}$ and ${\rm CeIn_3}$
the N\'{e}el temperature can be suppressed by applying
pressure\cite{Lonzarich}.
In the quantum critical regime, these materials indeed show
non-Fermi liquid properties.
The electrical resistivity is linear
(or close to being linear) in $T$, for as extended a temperature range
as three decades\cite{Steglich-lt23}.
The specific heat coefficient is either singular -- so that the 
effective mass diverges in the 
$T=0$ limit -- or is finite 
but contains a non-analytic dependence 
on temperature.
There is no doubt that such non-Fermi
liquid behavior originates from quantum critical physics, as Fermi
liquid properties (constant specific heat coefficient
and/or $T^2$ resistivity)
are recovered at low temperatures when the system 
is tuned away from
the QCP\cite{Gegenwart-prl02,Lohneysen,Grosche,Flouquet}.

Some
direct clues to the nature of such metallic QCPs
have come from the inelastic neutron scattering
experiments of Schr\"{o}der {\it et al.}\cite{Schroder,Stockert}.
The frequency and temperature dependences of the dynamical spin
susceptibility are characterized by an anomalous exponent $\alpha < 1$
as well as $\omega/T$ scaling. In addition, the 
same anomalous exponent $\alpha$ is seen essentially everywhere
in the Brillouin zone. These experimental results
differ in a very basic
fashion from the standard Hertz-Millis picture,
which argues for a Gaussian fixed point\cite{Sachdev-book}.
The Gaussian picture is formulated entirely in terms of
paramagnons -- the
long-wavelength fluctuations of the magnetic order parameter.
The critical theory is the $\phi^4$
theory, describing the non-linear couplings of the paramagnons.
Other degrees of freedom, including fermions, are considered to
be bystanders; the primary effect of fermions is to cause a Landau damping,
making the dynamic exponent, $z$, larger than 1.
In the antiferromagnetic case, the damping is linear in frequency and
$z=2$. For either three or two spatial dimensions ($d$)
the effective dimension, $d_{eff} = d +z$,
is larger than or equal to $4$, the upper critical dimension
of the $\phi^4$ theory.
The fixed point is therefore Gaussian. As a result, the dynamical spin
susceptibility would have to have the mean-field form,
\begin{eqnarray}
\chi^{Gaussian} ({\bf q},\omega) \sim
{1 \over {({\bf q} - { \bf Q})^2 - i \omega}} ,
\label{chi-rpa}
\end{eqnarray}
where ${ \bf Q}$ is the antiferromagnetic ordering wavevector.

The search for non-Gaussian quantum critical metals has
proceeded along a number of
directions\cite{lcp-nature,Coleman,Sachdev,Chubukov}.
Here, I describe in some detail the work on 
the local quantum critical
point\cite{lcp-nature,grempel-si,ZhuSi02,ZarandDemler02},
which can 
confront the
existing experiments.

\section{Model and microscopic approach}

\begin{figure}[t]
\begin{center}
\includegraphics[width=8.0cm]{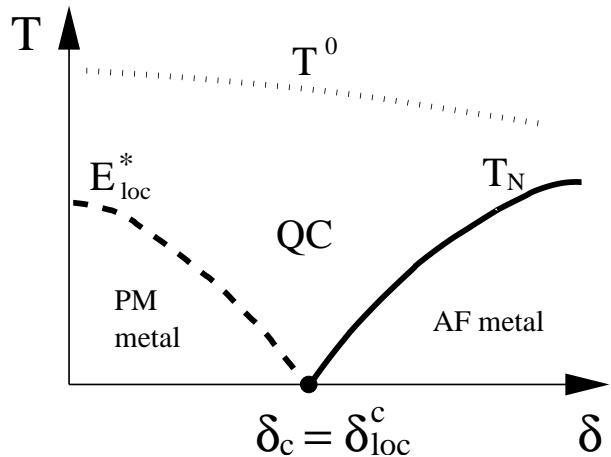}
\end{center}
\caption{
Local quantum critical point. 
The RKKY interaction $I \equiv T_K^0 ~\delta $ is tuned
as the bare Kondo scale $T_K^0$ is kept fixed.
The local susceptibility is 
Pauli below $E_{loc}^*$, Curie above 
$T^0$, and logarithmic [cf.\ Eq.\ (\ref{chi-loc-lqcp})]
in the quantum critical (QC) regime.
}
\label{lqcp}
\end{figure}

We focus on the Kondo lattice Hamiltonian,
\begin{eqnarray}
H
 = 
\sum_{ ij,\sigma} t_{ij} ~c_{i\sigma}^{\dagger}
~c_{j\sigma}
+ \sum_i J_{_K} ~{\bf S}_{i} \cdot {\bf s}_{c,i} 
+{1 \over 2}
\sum_{ ij} I_{ij} ~{\bf S}_{i} \cdot {\bf S}_{j} .
\label{kondo-lattice}
\end{eqnarray}
A lattice of spin-${1\over 2}$ local moments
(${\bf S}_i$) and a conduction electron band
($c_{i\sigma}$) are coupled through an antiferromagnetic 
Kondo exchange interaction ($J_{_K}$) and 
an RKKY interaction ($I_{ij}$).
The number of conduction electrons per unit cell, $x$, is 
taken to be close to but (without loss of generality)
less than 1; all the phases are metallic.
Two limits of this model are well-understood\cite{Doniach,Varma}.
When the RKKY interaction
is negligible, the local moments are expected to be completely screened
by the spins of the conduction electrons.
The resulting Kondo resonances
turn the local moments into
a part of the electronic excitations 
below some energy scale
$E_{loc}^*$ (Fig.~\ref{lqcp}).
In the opposite
limit, the dominating RKKY interactions lead to 
an antiferromagnetic metal.

To treat the dynamical competition between the RKKY and Kondo interactions,
we have applied an extended dynamical mean field theory 
(EDMFT)
\cite{SiSmith,ChitraKotliar00}.
The key quantities are the usual 
electron self-energy, $\Sigma(\omega)$, and a ``spin self-energy'',
$M(\omega)$, which enters the dynamical spin susceptibility
as follows\cite{SiSmith}
\begin{eqnarray}
\chi({\bf q},\omega) =
{1 \over {I_{\bf q} + M (\omega)}} ,
\label{chi-edmft}
\end{eqnarray}
where $I_{\bf q}$ is the spatial Fourier transform of $I_{ij}$.
The key
advantage of the approach is that here spin-damping is no
longer assumed to be 
due to 
a decay into 
quasiparticle-quasihole pairs. [Such an assumption is inherent 
in the paramagnon approach, and is responsible for a 
linear $\omega-$dependence in $M (\omega)$ -- 
cf.\ Eq.~(\ref{chi-rpa}).]
The approach sets out to also determine whether 
the fermionic excitations at the QCP
retain the heavy-quasiparticle character.
Within EDMFT, the lattice model is studied through a self-consistent
impurity model, the Bose-Fermi Kondo model:
\begin{eqnarray}
H_{\rm imp} = && 
J_K ~{\bf S} \cdot {\bf s}_c
~ + \sum_{p,\sigma} E_{p}~c_{p\sigma}^{\dagger}~ 
c_{p\sigma}
\nonumber\\
&& + \; g ~{\bf S} \cdot \sum_{p} 
\left( {\bf \phi}_{p} + {\bf \phi}_{-p}^{\;\dagger} \right)
+ 
\sum_{p} w_{p}\,{\bf \phi}_{p}^{\;\dagger}\cdot {\bf \phi}_{p} ,
\label{H-bfk}
\end{eqnarray}
where $g$, $E_p$ and $w_p$ are determined
self-consistently\cite{lcp-nature},
and 
${\bf \phi}_{p}$ describes a fluctuating magnetic field.

\section{Local quantum critical point}

We find two types of QCP\cite{lcp-nature}.
The more exotic type is the LQCP, as illustrated
in Fig.~\ref{lqcp}. Here, the local energy scale, $E_{loc}^*$, vanishes
at the QCP. The local Kondo physics is itself critical, and this
criticality 
is embedded in the criticality 
({\it i.e.} an infinite 
spatial 
correlation length)
associated with the magnetic ordering transition.

Corresponding to the vanishing $E_{loc}^*$ is a divergent local
($i.e.$, the ${\bf q}-$averaged) susceptibility. 
The specific form of the 
divergence is 
logarithmic,
\begin{eqnarray}        
  \chi_{{loc}} (\omega) ~= ~{ 1 \over {2 \Lambda}}
         ~\ln {\Lambda \over {-i \omega}} ,
\label{chi-loc-lqcp}
\end{eqnarray}
where $\Lambda \approx T_K^0$.
The spin self-energy is
\begin{eqnarray}        
 M (\omega ) ~\approx &&-I_{\bf Q} + A ~(-i \omega)^{\alpha} .
\label{M-lqcp}
\end{eqnarray}
Here, $I_{\bf Q}$ is the value of $I_{\bf q}$ at the ordering wavevector
${\bf q}={\bf Q}$, and the exponent is given by
\begin{eqnarray}        
\alpha = {1 \over {2 \rho_{_I}(I_{\bf Q}) \Lambda}} ,
\label{alpha-lqcp}
\end{eqnarray} 
where $\rho_{_I}(I_{\bf Q})$
is the ``RKKY density of states'',
$\rho_{_I}(\epsilon)  \equiv \sum_{\bf q} \delta 
( \epsilon - I_{\bf q})$ at $\epsilon =I_{\bf Q}$.

Viewed from the effective local Kondo model Eq.\ (\ref{H-bfk}),
the vanishing $E_{loc}^*$ corresponds to placing this local
problem on its critical manifold (a separatrix in the
$J_K-g$ parameter space). Indeed, 
the LQCP was first found\cite{lcp-nature}
using an $\epsilon \equiv 1 - \gamma$ 
expansion\cite{SmithSi97,Sengupta}
for Eq.\ (\ref{H-bfk}), along with the self-consistency
conditions.
Here $\gamma$ is the exponent that characterizes the spectrum of
the fluctuating magnetic field:
$\sum_{p} ~[ \delta (\omega - w _p) -
\delta (\omega + w _p) ]
 ~\sim ~|\omega|^{\gamma} ~{\rm sgn} \omega$.
The effects of spin-anisotropy 
($XY$ and Ising)
have been
treated in a similar fashion\cite{ZhuSi02,ZarandDemler02}.
More recently, numerical studies have been carried
out\cite{grempel-si} for the EDMFT equations in the anisotropic
Kondo lattice model,
using a Quantum Monte Carlo algorithm originally
developed by Grempel and Rozenberg\cite{grempel-rozenberg}.
The numerical results are consistent with the logarithmic
form for the singular local susceptibility at the
LQCP  [cf.\ Eq.\ (\ref{chi-loc-lqcp})].
The numerical value for the exponent $\alpha$ is about $0.7$, 
close to what is seen experimentally in ${\rm CeCu_{6-y}Au_y}$.

Since the peak susceptibility, $\chi ({\bf Q},\omega)$, 
is divergent at the QCP, the basic condition for realizing
a LQCP is such that the ${\bf q}-$averaged susceptibility
diverges at the same point
where the peak susceptibility does. Two dimensional magnetic fluctuations
satisfy this condition. If the magnetic fluctuations are purely
three dimensional, and if there is no frustration, then $E_{loc}^*$ 
would be finite at the QCP corresponding to
a crossover scale towards
the eventual Gaussian behavior\cite{lcp-nature,Burdin}.
We also note that our analysis applies provided 
the $T=0$ transition is continuous.
The latter can be explicitly checked by studying the EDMFT equations
on the ordered side 
(in the presence 
of 
a self-consistent static field);
work along this direction is in progress.

\section{Robustness of the LQCP}

Based on Ginzburg-Landau considerations\cite{lcp-nature},
we have also argued that
the LQCP is robust 
beyond the microscopic EDMFT provided $\alpha < 1$.
The 
key issue is whether the local susceptibility at the QCP
remains divergent when we allow for
a ${\bf q}-$dependence in the 
self-energies. Writing the general scaling form for the static
spin susceptibility for ${\bf q}$ in the vicinity of the ordering
wavevector ${\bf Q}$, 
\begin{eqnarray}
\chi({\bf q}) \sim { 1 \over
{ ({\bf q} - {\bf Q})^{2-\eta}}} ,
\label{chi-q-eta}
\end{eqnarray}
the question becomes equivalent to whether the spatial anomalous
dimension $\eta$ remains equal to zero: if it is, then the corresponding
local susceptibility in two dimensions remains singular.
For $\alpha < 1$, the non-linear couplings among the long-wavelength
modes are irrelevant; their contributions to the spin
self-energy will contain a ${\bf q}-$dependence that is at most 
$({\bf q} - {\bf Q})^2$. This, coupled with
the fact that the contributions to the spin self-energy from 
the local modes are expected to be smooth in ${\bf q}$, 
lead to the conclusion that $\eta =0 $. The LQCP is therefore
internally consistent.

To reiterate, 
the tuning to the QC regime provides a robust mechanism for
a singular local susceptibility, which in turn places the local
(Kondo) fluctuations 
exactly at 
its own criticality.

\section{Quantum-critical dynamics, Fermi surface properties, and experiments}

We now turn to
experimentally testable
properties:

\begin{itemize}

\item The dynamical spin susceptibility is 
 \begin{eqnarray}        
   \chi({\bf q}, \omega) = \frac{1}{
    (I_{\bf q} - I_{\bf Q}) + A ~(-i \omega )^{\alpha} 
           W (\omega/T) } .
\label{chi-qomegaT}
        \end{eqnarray}

\item The static uniform spin susceptibility
has a modified Curie-Weiss form,
        \begin{eqnarray}        
                \chi(T) = \frac{1}{        \Theta + B ~T^{\alpha}} ,
\label{modified-cw}
                \end{eqnarray}
with exactly the same exponent as in 
Eq.\ (\ref{chi-qomegaT}).

\item The NMR relaxation rate contains a temperature-independent component:
        \begin{eqnarray}        
        {1 \over T_1} ~\sim~ A_{hf}^2 ~{\pi \over { 8 \Lambda}} .
\label{1overT1}
        \end{eqnarray}

\item The Fermi surface 
changes sharply at the QCP, from ``large''
[volume $(1+x)$] to ``small''
(volume $x$ and a different topology) as the system orders.

\end{itemize}

Eq.~(\ref{chi-qomegaT}) reproduces the form observed\cite{Schroder}
in the inelastic neutron scattering experiments on
${\rm CeCu_{6-y}Au_y}$.
In addition, the 
neutron results
are consistent with two-dimensional magnetic
fluctuations\cite{Stockert}.

Recently, NMR experiments have been carried out\cite{Ishida} in
${\rm Yb Rh_2Si_2}$. The relaxation rate ${1 \over T_1}$ is strongly
dependent on the magnetic field, in a way that appears
consistent with $B/T$ scaling. Over the temperature range where 
the specific heat is logarithmic,
$\left ( {1 \over T_1} \right )_{B\rightarrow 0}$ 
does contain a constant component.

Even more recently, NQR experiments
have been carried out\cite{Walstedt} at a Cu site 
in ${\rm CeCu_{6-y}Au_y}$. The relaxation rate
has a non-Korringa temperature dependence;
at low temperatures ${1 \over T_1} ~\sim~ T^{\beta}$, 
where the exponent $\beta \approx 0.75$ is very close to
the fractional exponent $\alpha$ seen in the ${\bf q}$-dependent
dynamical spin susceptibility \cite{Schroder}.
This result would be consistent with the neutron
scattering result if the hyperfine couplings between
the Cu nuclei and the $f$-electron spins at the Ce sites 
are such that the dominant contributions to ${1 \over T_1}$ come
from generic wavevectors (wavevectors far away from the peak
wavevectors). With this assumption about 
the hyperfine coupling, the Cu-site NQR experiment confirms
the existence of the fractional exponent over an extended region
of the Brillouin zone.

In addition to ${\rm CeCu_{6-y}Au_y}$, the modified Curie-Weiss form for
the uniform spin susceptibility, Eq.~(\ref{modified-cw}), has also been
seen\cite{Steglich-lt23} in ${\rm Yb Rh_2Si_2}$.

The Fermi surface reconstruction can be probed through 
de Haas-van Alphen experiments.
We are aware of one case, ${\rm CeRh_2Si_2}$ (the quantum-critical
behavior of which is not yet as well characterized as for the other
heavy fermions), in which such measurements do 
reveal a Fermi-surface reconstruction\cite{Onuki}.
A less direct experimental signature is that the Hall coefficient 
should jump as the system is tuned through
the QCP\cite{JPCM,lcp-nature};
some preliminary indication for such a behavior has 
been found\cite{Paschen}
in ${\rm Yb Rh_2Si_2}$.

We make two remarks in passing. 
First, similar features in the dynamics
($\omega/T-$scaling)\cite{Aronson,MacLaughlin} and other
properties\cite{Maple,Stewart} occur in 
UCu$_{5-x}$Pd$_x$, raising the possibility that
local criticality is also operating
in these (strongly disordered and frustrated) systems.
Second, we have focused on the quantum-critical
physics associated with a transition between 
paramagnetic and antiferromagnetic metals.
Other phases, such as
spin-liquids\cite{Sachdev-Ye,Parcollet-Georges},
may also be relevant\cite{Sachdev}.

\section{Concluding remarks}

We have identified a local quantum critical metal, which is described by
a non-Gaussian fixed point. This picture provides a natural explanation
for some most unusual experiments in heavy fermion metals in the
vicinity of a magnetic quantum critical point.

The local quantum critical point serves as a concrete example in which
fermions, instead of being bystanders, directly participate in 
the critical behavior. The Kondo resonances, which have the 
quantum numbers of an electron and would turn
the local moments into a part of the electron fluid, become critical
at the same point where a magnetic ordering sets in.

More generally, the local moment physics in Kondo systems is analogous
to the Mott phenomenon.
The strong Coulomb interactions lead to a microscopic Coulomb-blockade
or, equivalently, a projection onto a restricted
Hilbert space.
Such effects are assumed to be inconsequential in the 
Gaussian quantum critical metal picture. On the other hand, and 
reminiscent of what happens in the usual Mott transition, they 
play a central role in a local quantum critical metal.


I would like to thank D.\ Grempel, K.\ Ingersent, 
S.\ Rabello, J.\ L.\ Smith, and L.\ Zhu for
collaborations, many colleagues -- particularly 
P.\ Coleman,
J.\ Custers,
J.\ Flouquet,
K.\ Ishida, 
H.\ v.\ L\"ohneysen,
G.\ Lonzarich, 
Y.\ \={O}nuki,
S.\ Paschen, 
C.\ P\'epin,
A.\ Rosch,
S.\ Sachdev,
F.\ Steglich,
and R.\ Walstedt
-- for discussions,
and NSF Grant No.\ DMR-0090071,
TcSAM, and the Robert A. Welch foundation 
for support.


\end{document}